# Electric Pulse Induced Resistive Switching, Electronic Phase Separation, and Possible Superconductivity in a Mott insulator

C. Vaju,[1] L. Cario,[1] B. Corraze,[1] E. Janod,[1] V. Dubost,[2] T. Cren,[2] D. Roditchev,[2] D. Braithwaite[3] and O. Chauvet[2]

[1] Institut des Matériaux Jean Rouxel (IMN), Université de Nantes, CNRS, 2 rue de la Houssinière, BP 32229, 44322 Nantes Cedex 3, France
[2] Institut des Nanosciences de Paris (INSP), CNRS UMR 75-88, Université Paris 6 (UPMC), 140 rue de Lourmel, 75015 Paris, France
[3] INAC, SPSMS, CEA Grenoble, 38054 Grenoble Cedex, France

Metal-insulator transitions (MIT) belong to a class of fascinating physical phenomena, which includes superconductivity,[1] and colossal magnetoresistance (CMR),[2] that are associated with drastic modifications of electrical resistance. In transition metal compounds, MIT are often related to the presence of strong electronic correlations that drive the system into a Mott insulator state. In these systems the MIT is usually tuned by electron doping or by applying an external pressure.[3] However, it was noted recently that a Mott insulator should also be sensitive to other external perturbations such as an electric field.[4,5] We report here the first experimental evidence of a non-volatile electric-pulse-induced insulator-to-metal transition and possible superconductivity in the Mott insulator $GaTa_4Se_8$. Our Scanning Tunneling Microscopy experiments show that this unconventional response of the system to short electric pulses arises from a nanometer scale Electronic Phase Separation (EPS) generated in the bulk material.

In the search for new phenomena that appear at the border between two different electronic states, the ternary chalcogenide $GaTa_4Se_8$ is a very attractive system to explore the frontier between the insulating and the metallic/superconducting states. $GaTa_4Se_8$ is a rare example of a stoichiometric Mott insulator that undergoes a pressure-induced insulator-to-metal and superconductor transition (at 11.5 GPa).[6,7] This is possible due to the relatively small value of the Mott-Hubbard gap (calculated to be around 0.1 eV),[6] by comparison with the large gaps (several eV) of usual inorganic Mott insulators.[3] The reasons for this must be sought in its particular crystal structure. $GaTa_4Se_8$ adopts a deficient spinel cubic structure of the $GaMo_4S_8$-type (see **Fig. 1**).[7,8] Compared to the regular spinel structure, the most interesting difference is a shift of the tantalum atoms off the centers of the chalcogenide octahedral sites, and the subsequent formation of tetrahedral $Ta_4$ clusters. The intracluster Ta-Ta distances are compatible with the formation of molecular bonds, while the larger intercluster distances prevent metal-metal bonding. This peculiar topology leads to the formation of molecular-like electronic states within the clusters and $GaTa_4Se_8$ is considered as a special class of Mott insulator in which the correlations take place not on the scale of single atoms but, instead, on the scale of small clusters.[6] This difference of scale reduces the magnitude of Coulomb repulsion and makes this compound easier to tune through the MIT.

Recently a new fascinating route to achieve the insulator-metal transition has emerged. Beside pressure and doping, theoretical studies have indeed shown that a sufficiently strong electric field could break the Mott insulator phase.[4,5] This could explain the volatile electric-field-induced MIT observed in systems such as Pr$_{1-x}$Ca$_x$MnO$_3$,[9] La$_{2-x}$Sr$_x$NiO$_4$,[10] or (LaS)$_{1.196}$VS$_2$.[11] Interestingly, (LaS)$_{1.196}$VS$_2$ and GaTa$_4$Se$_8$ bear some electronic and structural analogies : both compounds contain transition metal clusters and lie in the vicinity of the metal-insulator transition. This urges us to investigate the effects of electric pulses on the electrical transport properties of GaTa$_4$Se$_8$ crystals . As shown in Figure 2, the application of an electric pulse at 70K induces a non-volatile drop, of several orders of magnitude, of the electric resistance. Both curves of Figure 2a correspond to the low bias (lower than 0.01 V) resistivity $\rho(T)$ (resistance normalized to the form factor) measured with a standard four probes technique on the same crystal (sample #1). Curve 1 gives the resistivity of the pristine sample which shows an insulating behaviour with an average activation energy close to 0.07 eV. This behaviour changes radically after application of a 0.3 mA / 100 ms electric pulse, and the low bias After Electric Pulse (AEP) resistivity (curve 2) shows a drastic drop below 100 K. The same behavior was observed on a second sample (#2) measured by two-probe technique (Fig. 2b). For this sample, the MIT was initiated by a 1.8 mA/5 µs electric pulse at T=77 K . As shown in Figure 2b and f, the resistance of the sample during the pulse drops drastically (from 200 kΩ to 650 Ω) in less than 1 µs. The existence of the resistance switching, using our experimental setup, does not seem to critically depend on whether a current or a voltage pulse is applied. Figure 3 shows for another crystal (sample #3) immerged in liquid nitrogen the current-voltage characteristic measured by two-probe technique using 10 µs voltage pulses. At low positive voltage the resistance of the setup is 100 kΩ, and a linear variation consistent with the Ohm's law is observed. Suddenly, at 1.8 V the resistance of the sample drops drastically and the current reaches a high value (0.22 mA) just limited by the load resistor used in the circuit. This pulse leads to a non-volatile decrease of the setup resistance to 3 kΩ. This low-resistance state exhibits a linear dependence between -2V and +2V and therefore follows the Ohm's law. For this sample, the low resistance state is stable within this voltage range, but for some samples we observed that positive or negative voltage pulses could switch back the sample into its high-resistance state. A naive consideration of an homogeneous Joule heating cannot account for the non-volatile transition from high to low resistance : the energy $E_{Joule}$ dissipated during the short pulses is indeed very small ( $E_{Joule} = \int_0^{5\mu s} R_{sample} I^2 dt \approx 10$ nJ for sample #2, see Fig. 2 and 4 nJ for sample #3, see Fig. 3). This corresponds to a maximum temperature elevation of $\Delta T = 2$mK for a heat capacity of the crystal of 5 µJ/K (sample #2) estimated using the specific heat of other isostructural compounds,[12] and considering the worst hypothesis of thermally isolated samples. For all the samples investigated so far (more than 30), the resistance switching is observed whatever the temperature at which the pulses were applied (2K<T<100K). The number of pulses and the current density or electric field amplitudes required to induce the non-volatile drop of resistance are somewhat sample dependent, but an AEP state (like curves (2) or (4) in Figure 2a and b) is always generated. Therefore, all our observations reveal an exceptional behavior : application of short electric pulses can induce a non-volatile metal-insulator transition in the Mott insulator GaTa$_4$Se$_8$.

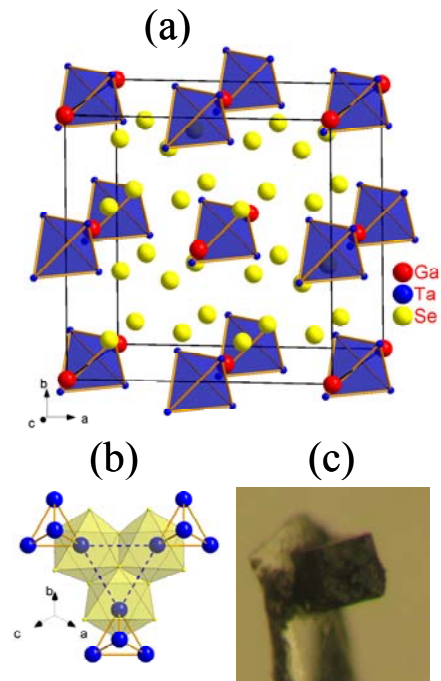

**Figure 1**. Structure of the chalcogenide spinel compound $GaTa_4Se_8$. (a) The cubic structure of $GaTa_4Se_8$ (space group F-43m) determined at 90K. (b) In this structure the tantalum atoms are shifted off the center of the chalcogenide octahedral sites which results in the subsequent formation of tetrahedral $Ta_4$ clusters ($d_{Ta-Ta}$ = 3.0050(4) Å, thick solid lines) separated by long Ta-Ta distances ($d_{inter\_Ta}$ = 4.3228(4) Å, dotted lines). (c) Micrograph of the single crystal used to perform the X-ray data collection on a Bruker CCD at 90K. The crystal was contacted with two carbon electrodes and electric pulses were subsequently applied. It was checked that the structure before the pulse ($R_{setup}$ = 190 kΩ; reliability factor $R_{obs}$ = 2.95%; cell parameter a= 10.3639(11) Å; $d_{Ta-Ta}$ = 3.0046(5) Å and $d_{inter\_Ta}$ = 4.3238(5) Å) is identical to the structure after the pulse ($R_{setup}$ = 120 kΩ; $R_{obs}$ = 2.71%; a= 10.3631(11) Å; $d_{Ta-Ta}$ = 3.0050(4) Å and $d_{inter\_Ta}$ = 4.3228(4) Å).

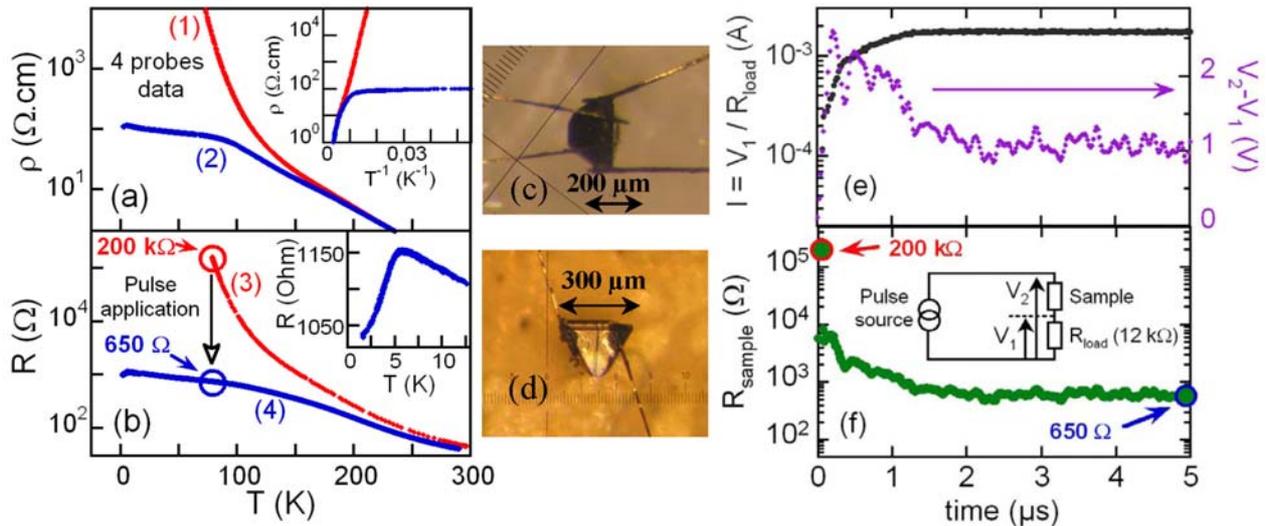

**Figure 2**. Resistive switching induced by electric pulses on GaTa$_4$Se$_8$ single crystals.
(a) Temperature dependence of the resistivity of a GaTa$_4$Se$_8$ crystal (sample #1) before (curve #1) and after (curve #2) a 0.3 mA/100 ms electric pulses. Data were obtained at low bias (typically 10 mV) on the crystal shown in (c) with a four probes technique. Inset: resistivity curves versus the reciprocal temperature.
(b) Low bias resistance of the GaTa$_4$Se$_8$ crystal (sample #2) shown in (d) measured using a two probes technique. Curves #3 and #4 were measured respectively before and after application of a 1.8mA/5µs pulse described in parts (e-f). Inset: zoom on the low temperature superconducting transition.
(c) GaTa$_4$Se$_8$ crystal (sample #1) used for measurement shown in part (a). The four gold wires were contacted using a carbon paste (Emerson & Cuming, Electrodag PR-406).
(d) GaTa$_4$Se$_8$ crystal (sample #2) corresponding to data shown in part (b). Gold electrodes separated by 10 microns were evaporated on the crystal. Gold wires were contacted on top of the gold electrodes using a carbon paste.
(e) Evolution of the current and voltage across the crystal (sample #2) during the 5 µs electric pulse .
(f) Resistance drop of sample #2 measured during the 5µs pulse of 1.8 mA. The resistance is calculated as $R_{sample} = (V_2 - V_1) / I = R_{load} (V_2 - V_1) / V_1$. Inset : schematic description of the measuring circuit.

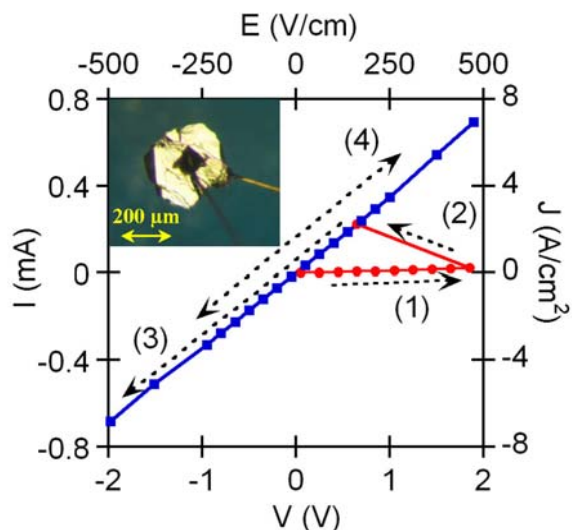

**Figure 3.** Current-Voltage characteristic of high resistance state (BEP in Red) and low resistance state (AEP in blue) measured with 10 µs voltage pulses on a crystal of GaTa$_4$Se$_8$ immerged in liquid nitrogen. The total Joule energy dissipated during the resistive switching is lower than 4 nJ. Inset : picture of the crystal connected in capacitance geometry.

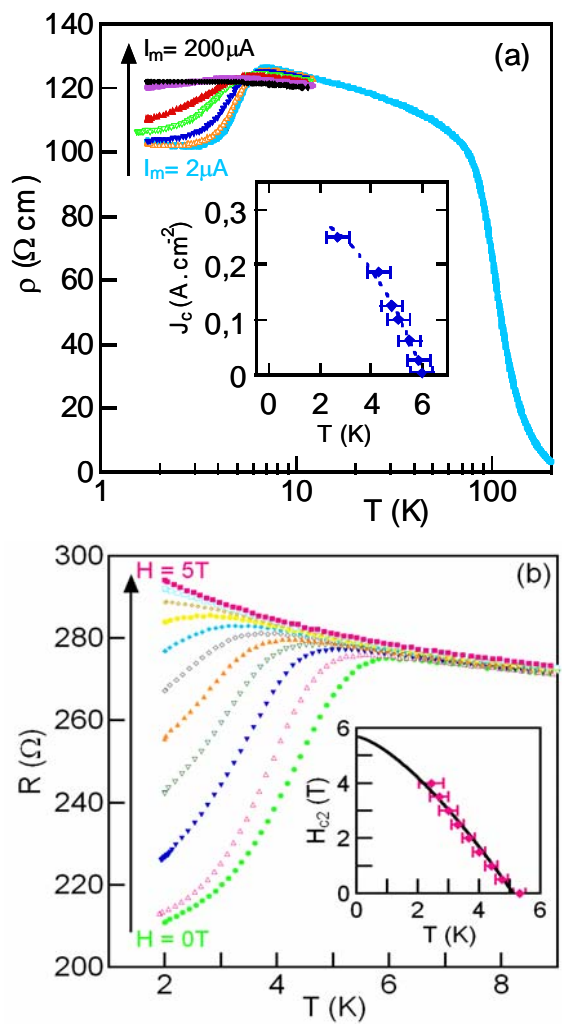

**Figure 4**. Critical current and critical field dependences of the resistivity (a) Low bias current evolution (from 2 to 200 µA) of the resistivity of sample #3 obtained with a 4 probes technique. The critical current density dependence of $T_C$ is shown in the inset. (b) Low bias resistance of sample #4 measured at different magnetic fields. Inset: temperature dependence of the upper critical field.

At lower temperature (< 10K), the "metallic-like" AEP state undergoes a further resistance drop below $T_C$ = 6 K (inset Fig. 2b). This transition can be easily suppressed by a short annealing of the sample (10 min at 250°C) and re-established by another electric pulse. We note moreover that the annealing only slightly increases the low-temperature resistivity and does not drive the system back to its initial high resistance state. The existence of this transition and a finite resistance below $T_C$ are perfectly reproducible from one crystal to another. In all cases we observed a relative variation $\Delta\rho/\rho$ of the resistivity ranging from 5 to 35 % at $T_C$ and the critical temperature $T_C$ in the 5-7 K range. To get some insights on the possible origin of the observed low temperature AEP transition, we studied the resistance response to the electric current and the magnetic field. This transition is gradually suppressed with increasing current density (Fig. 4a) or magnetic field (Fig. 4b), which might evoke a superconducting transition like the one observed in GaTa$_4$Se$_8$ under a pressure of 11.5 GPa at Tc = 5.8 K.[7] Nevertheless, no diamagnetic signal was revealed in our AC susceptibility experiments, within our resolution of 10$^{-7}$ emu. Interestingly, the (London) penetration depth of the magnetic field reaches $\lambda_L$ =1000 nm in the closely related superconducting Chevrel phases – which are also chalcogenides clustered compounds – with similar critical temperatures.[13] The apparent absence of diamagnetic Meissner and shielding effects might be easily explained by the complete penetration of the magnetic field within superconducting domains much smaller than $\lambda_L$. Therefore, all the above experiments indicate that the electric pulses have induced a non-percolating granular superconducting state in the Mott insulator GaTa$_4$Se$_8$. Electric-pulse-induced resistive switching was encountered in many compounds, but to our knowledge, this is the first time that such a striking behaviour has been observed.

The different mechanisms proposed for the resistive switching observed, so far, in transition metal oxides or chalcogenides, were recently reviewed :[14] they depend mainly on the class of materials and they can be classified according to whether the dominant contribution comes from a thermal effect, an interfacial electronic effect or an ionic effect. For example in the Cu- or Ag-based

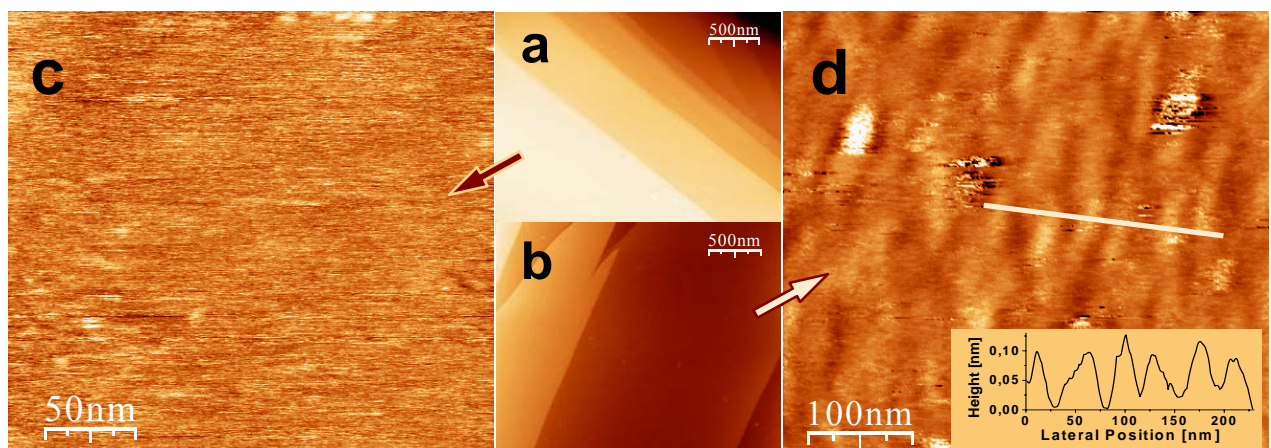

**Figure 5**. STM images of the surface of two cleaved $GaTa_4Se_8$ crystals. Large scale images of the pristine crystal in (a) (bias 0.76 V, IT = 0.25 nA) and of the transited crystal in (b) (bias 0.56 V, IT = 0.25 nA) reveal, in both cases, large terraces. On a nanometer scale however, clear differences are observed: While the pristine crystal surface shows no structure in (c) (bias 0.76 V, IT = 0.25 nA), the transited one displays a characteristic filamentary pattern in (d) (bias 0.44 V, IT = 0.17 nA). Inset in (d): The topography profile along the indicated line highlights the observed spatial modulation.

chalcogenides, like $Ag_2S$,[14] or in oxides like $SrTiO_3$,[15] the resistive switching is related to redox processes induced by cation or anion migration, and the mechanism refers mainly to an ionic effect. For the Ge- or Sb-based telluride alloys the resistive switching is related to the amorphous-crystalline phase transition observed in these systems,[16,17] and this effect can be classified as a thermal effect. Similarly, the resistive switch in NiO requires a high power density ($\approx 10^{12}$ W/cm$^3$)[14] and electric fields ($\approx 1$ MV/cm), and was associated to the formation or destruction of Ni conductive filaments by Joule heating.[18] Finally, in manganites,[19] or Cr-doped $SrTiO_3$,[20] the resistive switching mechanism refers mainly to an interfacial electronic effect. A charge doping model in which the charge injection induces an insulator metal transition close to the interface was initially proposed.[21] However, recent works on transition metal oxides have highlighted the role of the oxygen electromigration as another possible origin for the MIT observed close to the interface.[22, 23]

None of the mechanisms describe above seems appropriate to explain the resistive switching in the Mott insulator $GaTa_4Se_8$. For this compound, the resistive switching can be obtained with electric fields as low as 450 V/cm (sample #3, see Fig. 3) which is three orders of magnitude smaller than values required for the fuse type switching observed in NiO. Moreover, our single crystal X-ray diffraction experiments showed no changes, within our experimental resolution, in the structure of the crystal of $GaTa_4Se_8$ before and after application of pulses (see Fig. 1).[24] The resistive switching is thus neither related to a significant sample damage nor to drastic structural modifications such as the crystalline-amorphous transition observed in chalcogenides Phase Change Materials.[16, 17] The resistive switching in $GaTa_4Se_8$ was also observed using a four probes technique which is consistent with a "bulk" effect and not with an interfacial electronic effect as proposed in transition metal oxides like manganites.[21-23] Finally, contrary to Cu, Ag or O, none of the elements Ga, Ta and Se are known to lead to high ionic conductivity which makes redox processes induced by cation or anion migration very unlikely in $GaTa_4Se_8$. Similarly the capacitance geometry used for sample #3 (see inset of Fig.3) invalidates the scenario of a diffusion of electrodes elements (which would require a bulk diffusion of carbon atoms over more than 40 μm) for the origin of the resistance switching.

To get a deeper insight into the microscopic nature of the pulse-induced MIT observed in $GaTa_4Se_8$, STM measurements were performed. These experiments were undertaken on crystals cleaved in air just before their introduction into the ultra-high vacuum chamber (base pressure 2x10$^{-10}$ mbar). The STM images revealed a sample surface characterized by large flat terraces, observed on both, pristine (Fig. 5a) and transited (Fig. 5b) crystals. The well-defined steps clearly visible on both images have apparent height of 0.5 nm (or its multiple) corresponding to half the unit cell parameter of the material. The high-resolution STM images reveal, however, that important differences exist on a nanometer scale. Whereas the surface topography of pristine crystals is structure-less (Fig.5c), the STM images recorded on transited crystals are characterized by 30-50 nm size filamentary structures (Fig. 5d), with apparent height of ~0.1 nm, and qualitatively along the direction of the electric pulse. Such a filamentary pattern is representative of the AEP state: similar images were obtained with different STM tips in numerous locations of different transited crystals. These STM measurements indicate that the electric pulse affects somehow the bulk material as they were obtained on crystals cleaved in their AEP states. In constant-current STM images, regions with increased local density of states appear higher than regions with reduced local density of states. It is therefore not easy to distinguish *a priori* between real topographic structures and spectroscopic effects. However, an apparent height of ca. 0.1 nm lower than the atomic scale (ca. 0.2-0.3 nm) seems difficult to explain on the basis of pure topographic effects, and at least a part of the apparent height variations observed in Figure 5b has an electronic origin. Thus, STM images strongly suggest the presence of an Electronic Phase Separation (EPS) in the AEP state and point to the formation of nanometer-size domains organized in a filamentary structure. It is worth noting that EPS has recently attracted great attention in transition metal oxides that exhibit strong electronic correlations as $GaTa_4Se_8$.[25-30]

Considering such an electric pulse induced EPS, the peculiar transport properties observed in $GaTa_4Se_8$ (Fig. 2) can be naturally explained. In fact, the high temperature resistance is barely affected by the electric pulse while a strong reduction of the resistance occurs at low temperature in the AEP state (see inset Fig. 2a). This suggests that electric pulses generate a small amount of new conducting channels in parallel with unaffected parts of the crystal. Since the room temperature conductivity of the unaffected parts is rather high, a small number of filament-like conducting domains would indeed not change significantly the macroscopic resistance in this temperature range. Conversely, at lower temperature where the resistance of the unaffected parts increases drastically, the filamentary domains may become the leading conductance pathways and therefore correspond to the main contribution to the AEP resistivity below 100 K. The rather high value of the low temperature resistivity (ca. $10^2$ $\Omega$.cm, see curve 2 in Fig. 2a) compared to usual correlated metallic systems (typically smaller than $10^{-1}$ $\Omega$.cm),[3] is compatible with a small volume fraction of the conducting phase. Interestingly, the resistivity temperature dependence of this conducting phase resembles the one of granular metals. [31] This suggests that metallic domains are discontinuous and are separated by less conducting parts. This perfectly corresponds to the STM observation of nanometer size domains and reinforces the hypothesis according to which the resistance drop observed at $T_C$ = 5-7 K results from an inhomogeneous non-percolating superconductivity.

In conclusion, we have reported the first experimental evidence that an electric pulse can induce a non-volatile insulator-to-metal transition in the Mott insulator $GaTa_4Se_8$. Our results show that the breakdown of the Mott insulating state is concomitant with an electric-pulse-induced electronic phase separation. The peculiar metallic-like state exhibits an electronic transition at $T_C$ = 5-7 K which is presumably associated with the occurrence of a non percolative superconductivity. At this stage, a detailed microscopic picture of these phenomena is missing. The resistive switching observed in $GaTa_4Se_8$ seems quite different from the thermal, ionic, and electronic effects leading to resistive switching in other type of materials. We can not conclude yet on whether the resistive switching observed in $GaTa_4Se_8$ is a current or a voltage effect. We note, however, that $GaTa_4Se_8$ shows both, a pressure and an electric pulse induced MIT, and belongs to a non centrosymmetric Space Group (*F-43m*) compatible with piezoelectricity. Recent theoretical calculations have shown that the electric field could affect the transfer integral (*t*) in a Mott insulator and possibly lead to a MIT.[4,5] In principle, for piezoelectric compounds the transfer integral (*t*) could be even more affected by the electric field. Piezoelectric (or ferroelectric) effect could therefore provide a key to the understanding of the striking electric-pulse induced MIT an concomitant EPS formation observed in the Mott insulator $GaTa_4Se_8$.

**Experimental**

*Sample preparation :* Pure powders of $GaTa_4Se_8$ were synthesized as reported elsewhere [6-8]. Crystals of typical sizes around 300 μm were subsequently obtained by selenium transport method. The powder was placed in an evacuated silica tube with a small excess of Se and heated at 950°C for 24 hours and then slowly cooled (2°C/hour) to room temperature.

*Transport measurements :* The low bias resistance of the crystals was measured using a source measure Keithley 236 with a standard 2 or 4 probes technique and different types of electrodes (Au, Ag, C).

Electric pulse were applied using an Agilent 8114A. During the pulse the voltage drop in the sample was measured with a Tektronik TDS 340 oscilloscope, and the intensity through the circuit was monitored with a load resistor inserted in series.

*Description of the samples :* S: section of the sample; L: distance between injection electrodes; l: distance between bias measurement electrodes.
- Sample #1 shown in Figure 2c: four probes technique with carbon electrodes (Emerson & Cuming, Electrodag PR-406). L = 240 μm  l = 60 μm  S = 160 * 140 μm².
- Sample #2 shown in Figure 2d: two probes technique with gold electrodes evaporated on the crystal and separated by 10 microns. Gold wires were contacted on top of the gold electrodes using a carbon paste. L = l = 10 μm ; S = 200 * 100 μm²
- Sample #3 shown in Figure 3: cleaved crystal measured in capacitance geometry with two carbon electrodes (pad size = 100 * 100 μm²) L = l = 40 μm ; S= 200 * 200 μm² and .
- Sample #4 : four probes technique with carbon electrodes L = 300 μm; l = 60 μm; S = 260*200 μm².
- Sample #5 : two probes technique with carbon electrodes L = l = 140 μm ; S = 180 * 180 μm²

**Acknowledgments**
The authors thank J. Martial at IMN for her help in samples preparation, and F. Debontridder at INSP for technical support. This work was supported by a Young Researcher grant ANR-05-JCJC-0123-01 from the French Agence Nationale de la Recherche (to L.C., B.C. and E.J). L.C., B.C. and E.J contributed equally to this work. Supporting Information is available online from Wiley InterScience or from the authors.